\documentclass{article}

\pdfoutput=1
\usepackage{authblk}
\usepackage{epsfig}
\usepackage{epstopdf}
\usepackage{natbib}
\usepackage{fp}
\usepackage[capbesideposition=right]{floatrow}
\usepackage{wrapfig}
\usepackage[hmargin=1in,vmargin=1in]{geometry}

\newcommand{\Rs}{$ R_{\odot}$}
\newcommand{\Rf}{$ R_f$}

\def\ion[#1 #2]{#1\,{\sc #2}}

\begin{document}

\title{First Empirical Determination of the $\rm Fe^{10+}$ and $\rm Fe^{13+}$ \\
Freeze-in Distances in the Solar Corona}

\author[1]{Benjamin Boe}

\author[1]{Shadia Habbal}

\author[2]{Miloslav Druckm\"uller}

\author[3]{Enrico Landi}

\author[1]{Ehsan Kourkchi}

\author[4]{Adalbert Ding}

\author[2]{Pavel Starha}

\author[5]{Joseph Hutton}

\affil[1]{\small Institute for Astronomy, University of Hawaii, Honolulu, HI 96822, USA}
\affil[2]{\small Faculty of Mechanical Engineering, Brno University of Technology, Technick 2, 616 69 Brno, Czech Republic}
\affil[3]{\small Department of Climate and Space Sciences and Engineering, University of Michigan, Ann Arbor, MI 48109, USA}
\affil[4]{\small Institute of Optics and Atomic Physics, Technische Universitaet Berlin, Germany}
\affil[5]{\small Aberystwyth University, Wales, UK}

\maketitle

\begin{abstract}
Heavy ions are markers of the physical processes responsible for the density and temperature distribution throughout the fine scale magnetic structures that define the shape of the solar corona. One of their properties, whose empirical determination has remained elusive, is the `freeze-in' distance (\Rf) where they reach fixed ionization states that are adhered to during their expansion with the solar wind. 
We present the first empirical inference of \Rf\ for Fe$^{10^+}$ and Fe$^{13^+}$ derived from multi-wavelength imaging observations of the corresponding \ion[Fe xi]  (Fe$^{10^+}$) 789.2 nm and \ion[Fe xiv] (Fe$^{13^+}$) 530.3 nm emission acquired during the 2015 March 20 total solar eclipse. We find that the two ions freeze-in at different heliocentric distances. In polar coronal holes \Rf\ is around 1.45 \Rs\ for Fe$^{10^+}$ and below 1.25 \Rs\ for Fe$^{13^+}$. Along open field lines in streamer regions \Rf\ ranges from 1.4 to 2 \Rs\ for Fe$^{10^+}$ and from 1.5 to 2.2 \Rs\ for Fe$^{13^+}$. These first empirical \Rf\ values: (1) reflect the differing plasma parameters between coronal holes and streamers and structures within them, including prominences and Coronal Mass Ejections (CMEs); (2) are well below the currently quoted values derived from empirical model studies; and (3) place doubt on the reliability of plasma diagnostics based on the assumption of ionization equilibrium beyond 1.2 \Rs.
\end{abstract}

\section{Introduction} 
\label{intro}

During the total solar eclipse of 1869 August 7, Young and Harkness  \citep{Young1870, Young1871, Young1872} discovered a bright line in the coronal spectrum, hitherto unknown. The work of \cite{Grotrian1934, Grotrian1939} and \cite{Edlen1945} 
subsequently led to its identification as emission from Fe$^{13^+}$ (Fe XIV) at 530.3 nm. The presence of forbidden emission from such a highly ionized state of iron indicated that the solar corona was at temperatures exceeding a million degrees. Over a decade later, \cite{Parker1958} demonstrated with a simple isothermal model that such a hot atmosphere could not remain bound to the Sun and should therefore expand in an outflow, which he coined the `Solar Wind'. The first unquestionable detection of a flux of positive ions of solar origin (i.e. the solar wind) was made by the Mariner 2 spacecraft 
\citep{Neugebauer1966} sent to Venus in late 1962. This detection followed those of the earlier Russian probes launched between 1959 and 1961 \citep{Gringauz1960}, and the American Explorer 10 in 1961 \citep{Bonetti1963}, which were less conclusive given that these spacecraft did not always clear the magnetosphere. Mariner 2 detected a continuous plasma flow with speeds ranging from 300 to 800 km s$^{-1}$ and temperatures averaging $1.7 \times 10^5$ K.

These early measurements were confirmed by subsequent in-situ probes such as Helios (e.g. \citealt{Marsch1983}) and Ulysses (e.g. \citealt{McComas1998}), to name a few. 

\par

The launch of space-based telescopes to observe the corona in the extreme ultraviolet (EUV) such as from the HCO experiment on OSO-4 (e.g., \citealt{Reeves1970}) and ATM on Skylab \citep{Reeves1972, Huber1974}, demonstrated that the solar corona was highly structured by magnetic fields with the brightest EUV emission originating from active regions. The darkest regions were dubbed coronal holes that were later shown by \cite{Munro1972} to have a significantly reduced density and temperature in comparison to their surroundings.
Imaging the corona in the EUV has the advantage of `seeing' the corona as projected on the solar disk with exquisite detail, as demonstrated by the more recent Atmospheric Imaging Assembly (AIA) instrument on the Solar Dynamic Observatory \citep{SDO2012}. However, the extension of the EUV emission off the solar limb remains limited to heliocentric distances of 1.5 \Rs\ at most. 

\par

With the exception of total solar eclipses, whereby emission is detected out to at least 2 \Rs, imaging in the visible wavelength range requires the use of coronagraphs as first demonstrated by \cite{Lyot1932} and implemented on SOHO/LASCO C2, C3 and STEREO/COR1 and COR2 \citep{SOHO1995, STEREO2008}. At present, ground- and space-based coronagraphs are limited by the size of their occulter, which is significantly larger than the angular size of the solar disk, as well as by diffraction effects at their edge. These limitations thus prevent them from imaging the corona down to the solar surface. Furthermore,  space-borne coronagraphs operating at present are limited to white light imaging with no spectroscopic diagnostics for measuring coronal emission lines. Consequently there exist both spatial and spectral gaps in the data currently available from the suite of EUV and white light observations. 

\par

Despite their paucity and short duration, total solar eclipses, at present, remain the only observational opportunities where imaging of the corona can be achieved in an uninterrupted manner from the solar surface out to several \Rs. These observations thus cover the distance range where the most rapid changes in the dynamics and thermodynamics of coronal structures occur, a distance range that cannot be covered by other remote sensing techniques at present. These attributes are essential for tracing the solar wind streams from their detection in-situ back to their sources at the Sun.

\par

Attempts to coordinate remote sensing with in-situ observations led to the association of the fastest streams with coronal holes \citep{Krieger1973}. On the other hand, the source regions of the slower streams, with speeds below 400 km s$^{-1}$, continue to be the subject of a plethora of investigations.Some have proposed closed magnetic regions for the origin of the slow wind, which   would release their plasma content into the solar wind as a consequence of magnetic reconnection (e.g. \citealt{Stakhiv2015}; \citealt{Stakhiv2016}). Others invoke open field lines at the boundaries of streamers (e.g. \citealt{Antiochos2011}; \citealt{Riley2012}) and/or active regions and coronal holes (e.g. \citealt{Sakao2007}, \citealt{Stakhiv2016}). Pseudostreamers have also been considered (e.g. \citealt{Wang1990}). The complexity and range of the plasma parameters measured in solar wind streams likely imply that there are in fact a multitude of different sources \citep{Xu2015}. 
\par

One of the promising approaches for establishing the link between in-situ measurements and the source regions of solar wind streams at the Sun is to explore the charge state evolution of heavy ions in the corona. Although heavy ions constitute only a very small fraction of the solar wind plasma, they 
are important indicators of the plasma conditions in the inner corona that define the ion charge states in interplanetary space (\citealt{Hundhausen1968}; \citealt{Owocki1983b}; \citealt{Ko1997}; \citealt{Landi2012b} and \citealt{Gloeckler2007}).

\par

Since the densities of electrons, protons and heavy ions decrease with heliocentric distance, the relative abundance of ionized species will vary until the electron density becomes low enough to prevent them from further ionization and recombination at some distance from the Sun. This distance where ionized species cease to change, and thus freeze into their respective ionization states, is called the `freeze-in' distance (\Rf) \citep{Hundhausen1968}. Charge states measured in situ therefore reflect the plasma conditions below \Rf\ in the corona. One approach to link in-situ plasma properties (i.e. species density, temperature and flow speed) to the physical conditions in the inner corona, in the absence of a direct measure of \Rf, is to resort to model studies. Such models use in-situ charge state measurements as empirical constraints to calculate the evolution of the plasma ionization, electron temperature and density, and bulk speed below \Rf\ (\citealt{Owocki1983b}; \citealt{Ko1997} and \citealt{Landi2012b}). However, these empirical models have yet to account for the complexity of coronal magnetic structures, in particular in streamers where eclipse observations make it amply evident that open field lines are present in addition to large scale loops (see Fig. \ref{fig1}).

\par
In this paper we show how observations of coronal emission in optical forbidden lines during total solar eclipses offer a novel empirical tool to infer \Rf\ for different ions. This empirical tool overcomes the limitations of models, which have been the only technique available so far to establish this link. The concept was originally discovered and demonstrated by \cite{Habbal2007} (see also \citealt{Habbal2013}). It is implemented here in a comprehensive manner for the first time, using the 2015 March 20 eclipse observations of two charge states of Iron: Fe$^{10^+}$ and Fe$^{13^+}$ (see lower panels in Fig. \ref{fig1}). We first give a brief description of the total solar eclipse observations (Section \ref{eclipse}),  followed by the details of the technique used to infer the freeze-in distance (Section \ref{sec_FreezeIn}). The results (Section \ref{sec:res}) illustrate how the complexity of coronal structures cause fluctuations in \Rf\ as discussed at length in Section \ref{sec:disc} and summarized in Section \ref{sec:conc}.


\section{Eclipse Observations}
\label{eclipse}

The 2015 March 20 total solar eclipse observations presented here were taken from the island of Svalbard in Norway at the hangar of the Longyearbyen airport, which was at N78$^o$14'48.6", E15$^o$29'21.3" with an altitude of 15 m. The Sun was 11$^o$ above the horizon during totality, which occurred between 10:10:40 and 10:13:08 UT under clear sky conditions. 

\par
\begin{figure*}[h!]
\centering
\includegraphics[width = 4.9in]{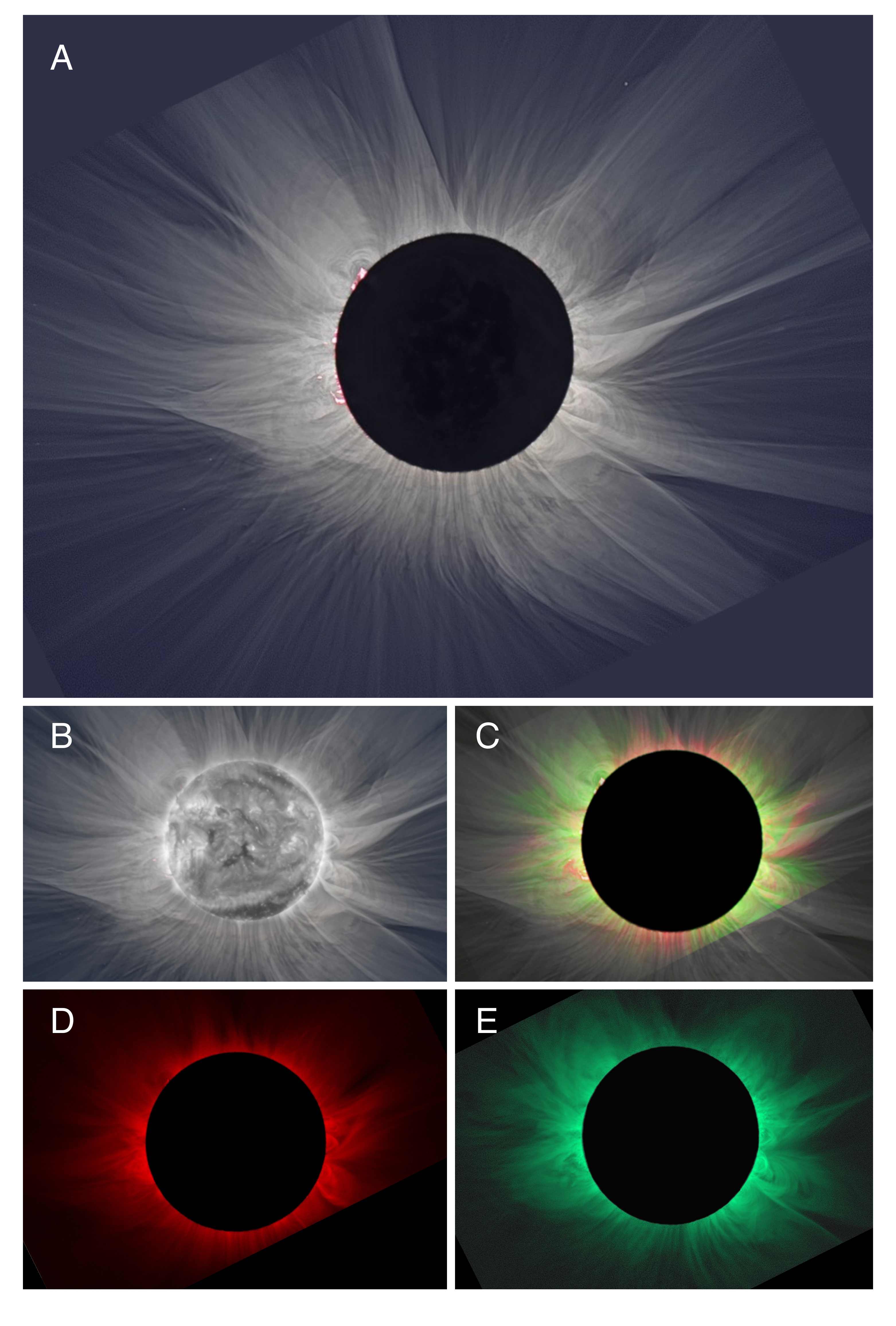}
\caption{Total solar eclipse observations from 2015 March 20. Solar north is vertically up in all images. A: White light image of the solar corona. B: Overlay of the white light image with an SDO/AIA 19.1 nm filter image taken within the time frame of totality. C: Composite \ion[Fe xi] 789.2 nm emission at $1.1 \times 10^6$ K (red) and \ion[Fe xiv] 530.3 nm emission at $1.8 \times 10^6$ K (green) overlaid on the white light image. Due to an inadvertent off-center pointing of the \ion[Fe xi] system during totality, part of the corona in the southwest is missing in comparison with \ion[Fe xiv]. D: \ion[Fe xi] emission. E: \ion[Fe xiv] emission.  }

\label{fig1}
\end{figure*}

The broad band white light image shown in panel A of Figure \ref{fig1} is a composite of 29 images taken with a Nikon D810 camera, retrofitted with a 115 mm aperture TS photoline apochromatic triplet with a focal length of 800 mm and a 2.5 inch field flattener. A sequence of exposure times ranging from 1/1600 s to 4 s was repeated throughout totality. The 29 images thus obtained were calibrated by means of dark frames and flat-fields. They were aligned by means of phase correlation and combined using the LDIC 5.0 software. They were then processed using the ACC 6.1 software based on the techniques developed by M. Druckm\"uller in order to visualize coronal structures
\citep{Druckmuller2006, Druckmuller2009}.  A composite of this white light image with an SDO/AIA 19.1 nm image taken at the same time as the eclipse (superimposed on the darkened solar disk) is shown in Figure \ref{fig1}B.There was a large coronal hole in the south and a smaller one in the north whose extensions are readily captured in this composite eclipse image.  
 
Images of emission from the \ion[Fe xi] 789.2 nm (i.e. Fe$^{10^+}$) and \ion[Fe xiv], 530.3 nm (i.e. Fe$^{13^+}$) coronal forbidden lines, with peak ionization temperatures of $1.1\times 10^6$ and $1.8 \times 10^6$ K respectively, were obtained with different optical systems. These systems consisted of 3" diameter, 300 mm focal length achromats retrofitted with 0.5 nm bandpass filters manufactured by Andover corporation. Data were recorded with Atik 314L monochrome cameras. Given that photons resulting from Thompson scattering of the photospheric radiation by coronal electrons contribute to the emission in each bandpass, additional systems were required for each spectral line to measure the background continuum. For each spectral line observation, one filter was centered on the wavelength of the emission line, while the other was centered 1 nm towards the blue from line center, namely at 788.2 and 529.2 nm for \ion[Fe xi] and \ion[Fe xiv] respectively. These continuum observations are referred to as `offband' images in what follows. The sequence of exposure times for these systems ranged from 0.3 to 6 seconds. These were then combined to provide a dynamic range sufficient to record the coronal emission from the solar surface out to the edge of the field of view enabled by the optics. Standard dark frame removal and flat fielding were applied to the images. The offband images were then subtracted from the spectral line centered images to isolate the emission from the desired spectral lines.

\begin{figure*}[t!]
\centering
\includegraphics[width = 5.6in]{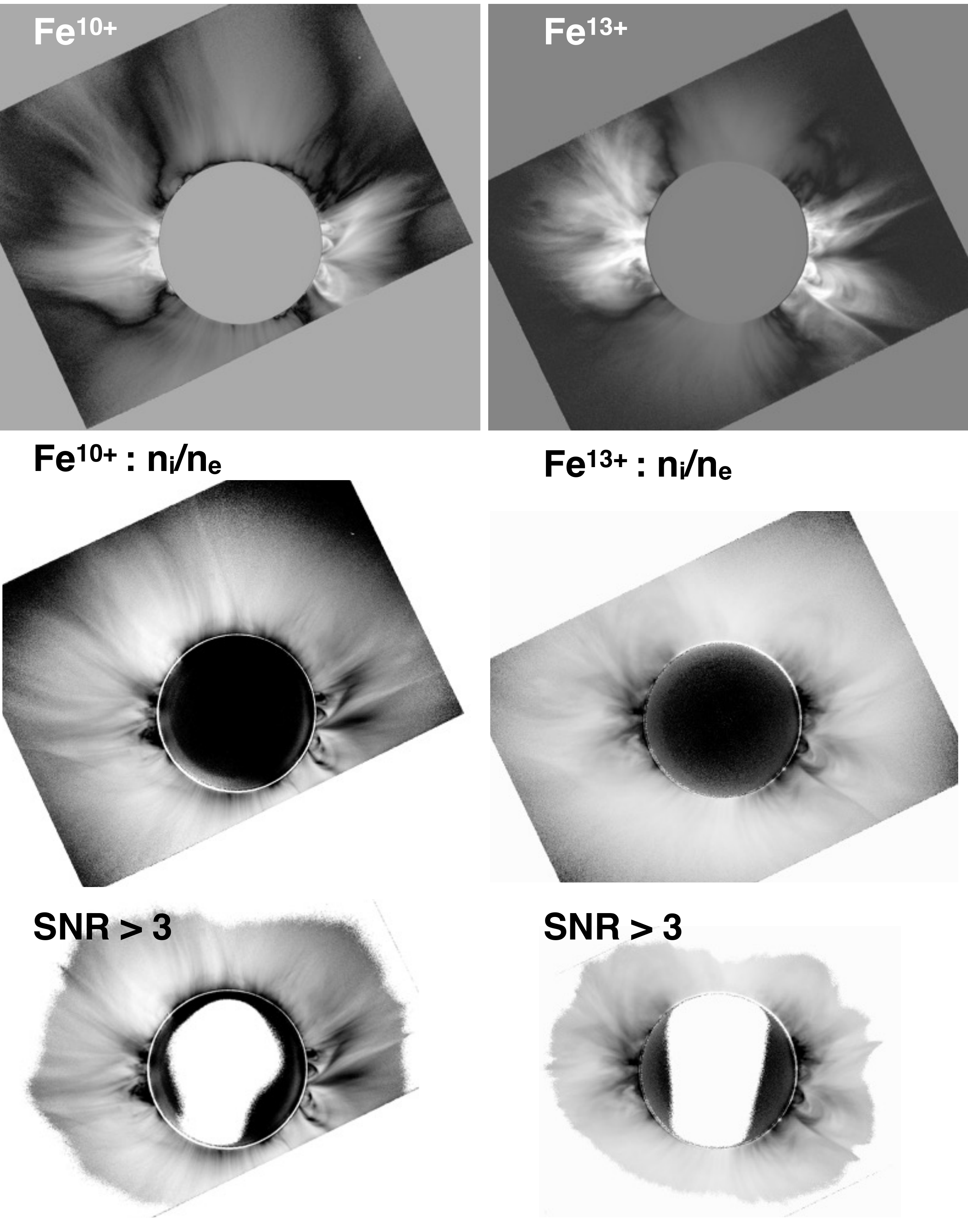}
\caption{Images of the narrowband data used throughout this work. The top panels shows the onband images for both \ion[Fe xi] (left) and \ion[Fe xiv] (right) which have been flattened by dividing out the exponential drop off in brightness (the brightness in the image shows deviations from the average brightness at that solar distance). The middle panels show the onband/offband ratio, which is $\approx n_i/n_e$, for \ion[Fe xi] (left) and \ion[Fe xiv] (right). Black and white are inverted in these panels to show detail, so a higher ratio value is darker. The bottom images (also color inverted) show the ratio images where the signal-to-noise ratio (SNR) was $>$ 3 for both onband and offband data. The noise level was taken from measurements at the center of the Moon. 
}
\label{fig2}
\end{figure*}

The resulting images of isolated forbidden line emission (referred to as `onband' images in what follows) are shown in Figure \ref{fig1}D and E. Unfortunately, the pointing of the \ion[Fe xi] system was not properly centered in the field of view during totality, causing some of the emission to be cut-off in a quadrant extending from the south to the west. A color composite of the \ion[Fe xi] (red), \ion[Fe xiv] (green) and white light images is shown in \ref{fig1}C, where the color balance was determined by the relative intensities of each emission line. A comparison of the two shows how different coronal structures appear in each wavelength. Indeed, the \ion[Fe xi]/\ion[Fe xiv] composite shows the dominance of \ion[Fe xi] emission in the coronal holes while \ion[Fe xiv] emission is more prevalent in streamers. The top panels of Figure \ref{fig2} show both \ion[Fe xi] and \ion[Fe xiv] onband images which have been flattened by dividing out the exponential drop in intensity. The equatorial region dominated by streamers shows a higher emission than the coronal holes for both ions, indicating a higher bulk density in the streamers than the holes. The darkest region traces out a distinct boundary between the polar holes and equatorial regions. 
\par
We note that the 2015 total solar eclipse coincided with the declining phase of activity in solar cycle 24 and had a white light corona `shape' characteristic of such periods (see Figure \ref{fig1}A).  Eclipse images taken in these two spectral lines at solar minimum (see 2008 eclipse image in \citealt{Habbal2010}) show a clear distinction between coronal holes and streamers, with the latter dominated by \ion[Fe xiv] emission within their bulges formed by closed large-scale loops. \ion[Fe xi] emission is also present in the streamers however, and becomes dominant in pockets of open field lines seemingly intermixed throughout the streamers. The presence of \ion[Fe xi] within the streamer regions in 2015, and likewise during solar minimum, suggests the presence of open field lines along the lines of sight intercepting streamers. Bound loops are typically heated much beyond the peak ionization temperature of \ion[Fe xi] ($1.1 \times 10^6$ K) and closer to that of \ion[Fe xiv] ($1.8 \times 10^6$ K), therefore the presence of the colder plasma indicates that some open field lines are intermingled around the bound lines, allowing the colder plasma to exist near the streamers (see Fig \ref{fig1}C). Furthermore, it is clear from the white light corona in Figure \ref{fig1}A that there are open field lines emanating from every direction from the Sun, which can be traced back to the top of bound loops, if not all the way to the solar surface.  

\section{Technical Approach}
\label{sec_FreezeIn}

\cite{Habbal2007, Habbal2013} were the first to show that measurements of emission from coronal forbidden lines and their neighboring continuum provide an empirical tool for the inference of the ion freeze-in distance (\Rf). They developed a technique based on the processes that lead to forbidden line emission.  The intensity of emission from highly ionized Fe$^{i+}$, where $i$ is the ionization state, can be separated into two main processes: collisions with electrons and radiative excitation by solar disk radiation, or: 
\begin{eqnarray}
I_{line} & = & I_{coll} + I_{rad}
\end{eqnarray}

\noindent
The first term $I_{coll}$ is proportional to the product of the density of the emitting ion $n_i$, in a given ionization state $i$, and the density of the exciting free electrons, $n_e$, which can be written as 

\begin{eqnarray}
I_{coll} & \propto  & n_i \times n_e 
\end{eqnarray}

\noindent
It is important to note that the value of $n_i$ is determined by the interplay between collisional ionization of Fe$^{(i-1)+}$ and dielectronic recombination of Fe$^{(i+1)+}$ with free electrons; these processes are proportional to $n_{i-1} \times n_e$ and $n_{i+1} \times n_e$ respectively. (Here we have ignored simple recombination and collisonal electronic excitation because their cross-sections are too small. Charge exchange with protons is also negligible due to the low proton energy). Radiative excitation, $I_{rad}$, is proportional to the number density of ions $n_i$ in ionization state $i$, i.e.

\begin{eqnarray}
 I_{rad} & \propto &  n_i 
\end{eqnarray}

 \noindent
 The intensity of the continuum radiation $I_{cont}$, due to Thompson scattering of photospheric radiation by electrons, is proportional to $n_e$, i.e.
 
 \begin{eqnarray}
 I_{cont} & \propto &  n_e 
\end{eqnarray}

\noindent
The ratio of line to continuum emission (onband/offband) is then proportional to 
$ n_i + n_i/n_e$. 
Since the density decreases faster than exponentially with radial distance, the second term $n_i/n_e$ (due to radiative excitation) quickly becomes the dominant one. The ratio of line to continuum can then be written as:
\begin{eqnarray}
\frac {n_i} {n_e} & = & \frac{n_i}{n_x} \times \frac{n_x}{n_p} \times \frac{n_p}{n_e} 
\end{eqnarray}
where $x$ refers to Fe. The last term is the proton to electron density ratio, which, in an almost fully ionized plasma such as the solar wind, is constant. The second term is the elemental abundance, which is fixed. Hence $\frac {n_i} {n_e}$ varies with distance like $\frac{n_i}{n_x}$. At the radial distance \Rf, when ionization and recombination stop, $\frac{n_i}{n_x}$ becomes constant and so does $\frac {n_i} {n_e}$. The freeze-in distance can therefore be measured as the distance along a field line where this ratio becomes constant.

\par
Taking the ratio of the onband and offband image intensities along the same radial scan in both images yields a direct tool for the inference of \Rf\ with eclipse observations. Such radial scans must be made along an `open' field line structure: i.e. one that extends outward from the Sun uninterrupted by bound structures. In regions where the structures visibly belong to closed loops, plasma is not actively escaping the corona and so will not have an \Rf. The measurements are therefore carried out only above the closed structures when the field lines become open. 
\par
The application of this concept to the eclipse images is as follows: Hand-selected points along a given `open' structure are interpolated in polar coordinates; the ratio is then computed between the onband and offband images for every pixel crossing the interpolated line; the resulting ratio profile is used to infer \Rf\ as the location where the profile flattens out indicating a constant ion density with respect to the electrons. The procedure is repeated for over one hundred separate lines around the Sun.

\begin{figure*}[h!]
\centering
{\includegraphics[width=5.4in]{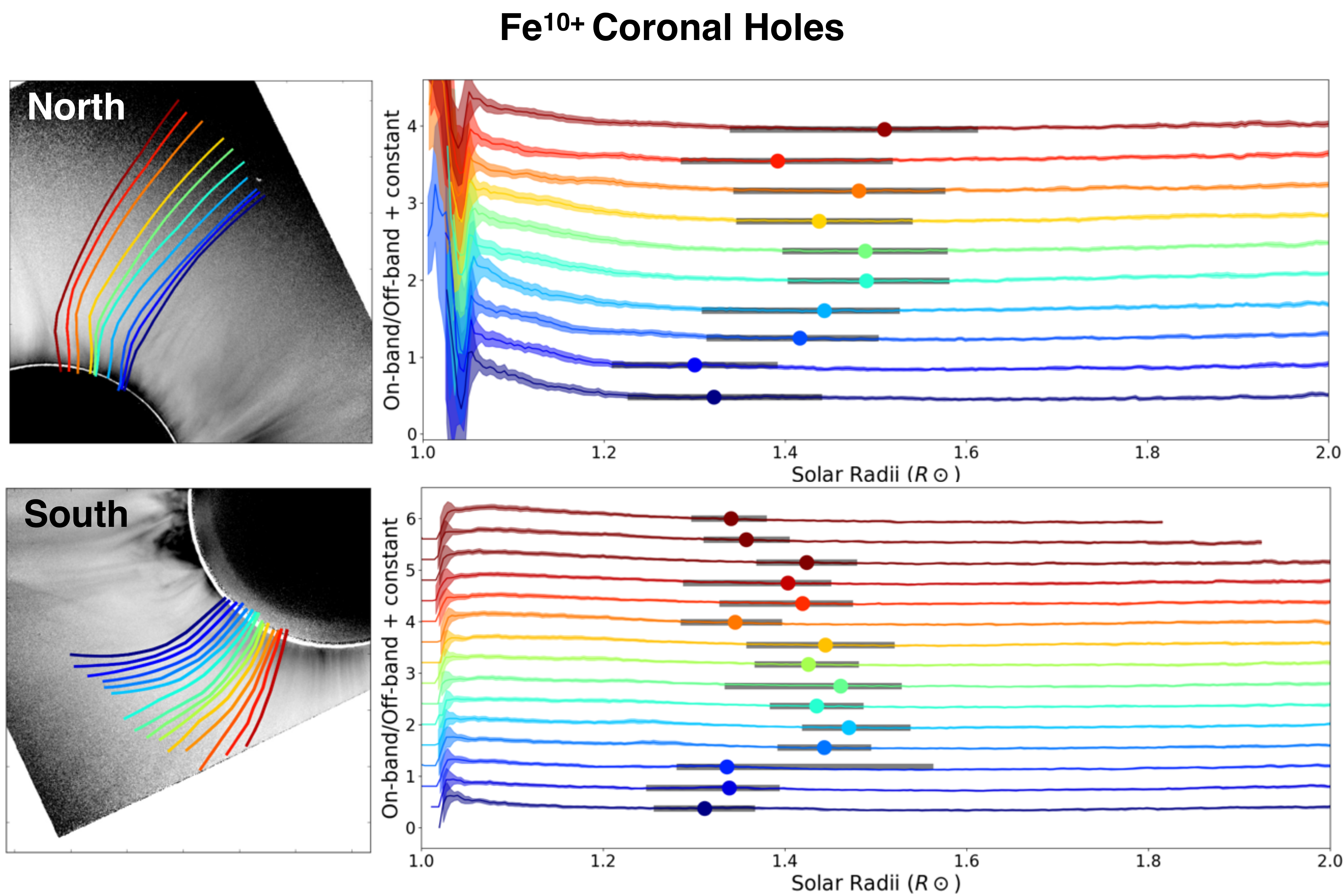}}\\
\vspace*{3mm}
\centering
{\includegraphics[width=5.4in]{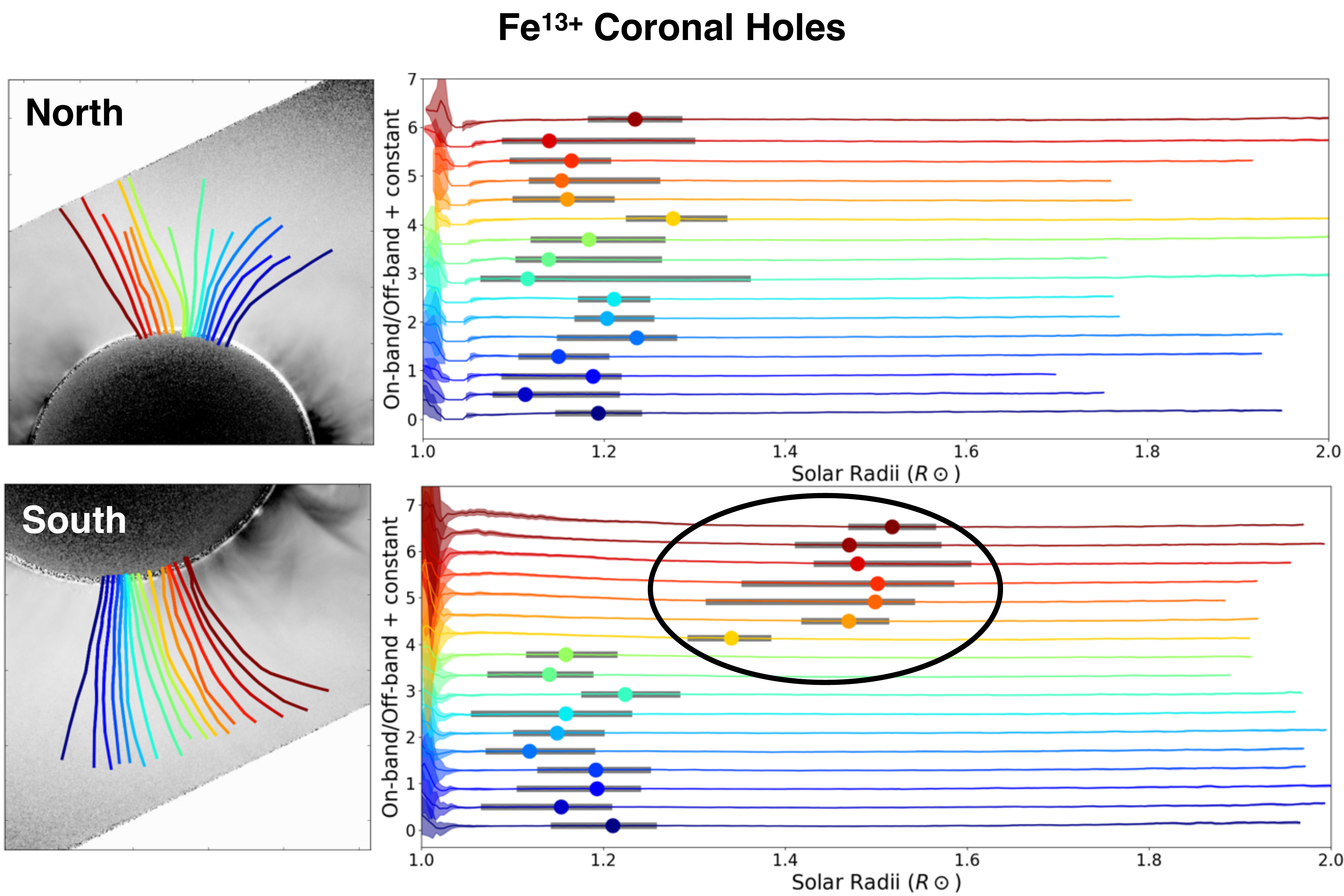}}
\caption{Example of drawn lines in the polar coronal holes in the north and south, for $\rm Fe^{10+}$ (top) and $\rm Fe^{13+}$ (bottom). The left panels show the drawn lines superposed on the ratio images from Figure \ref{fig2}. The right panels show the value of the onband/offband ratio along the drawn line (with a constant offset for each successive line). \Rf\ is given by the colored dots with error bars determined using the technique outlined in Section \ref{sec_FreezeIn}. Note that the dots encircled by an ellipse in the $\rm Fe^{13+}$ panel correspond to \Rf\ values measured around the boundaries between the coronal hole and the streamer to its north.}
\label{fig3}
\end{figure*}

\begin{figure*}[h!]
\centering
\includegraphics[width=6.0in]{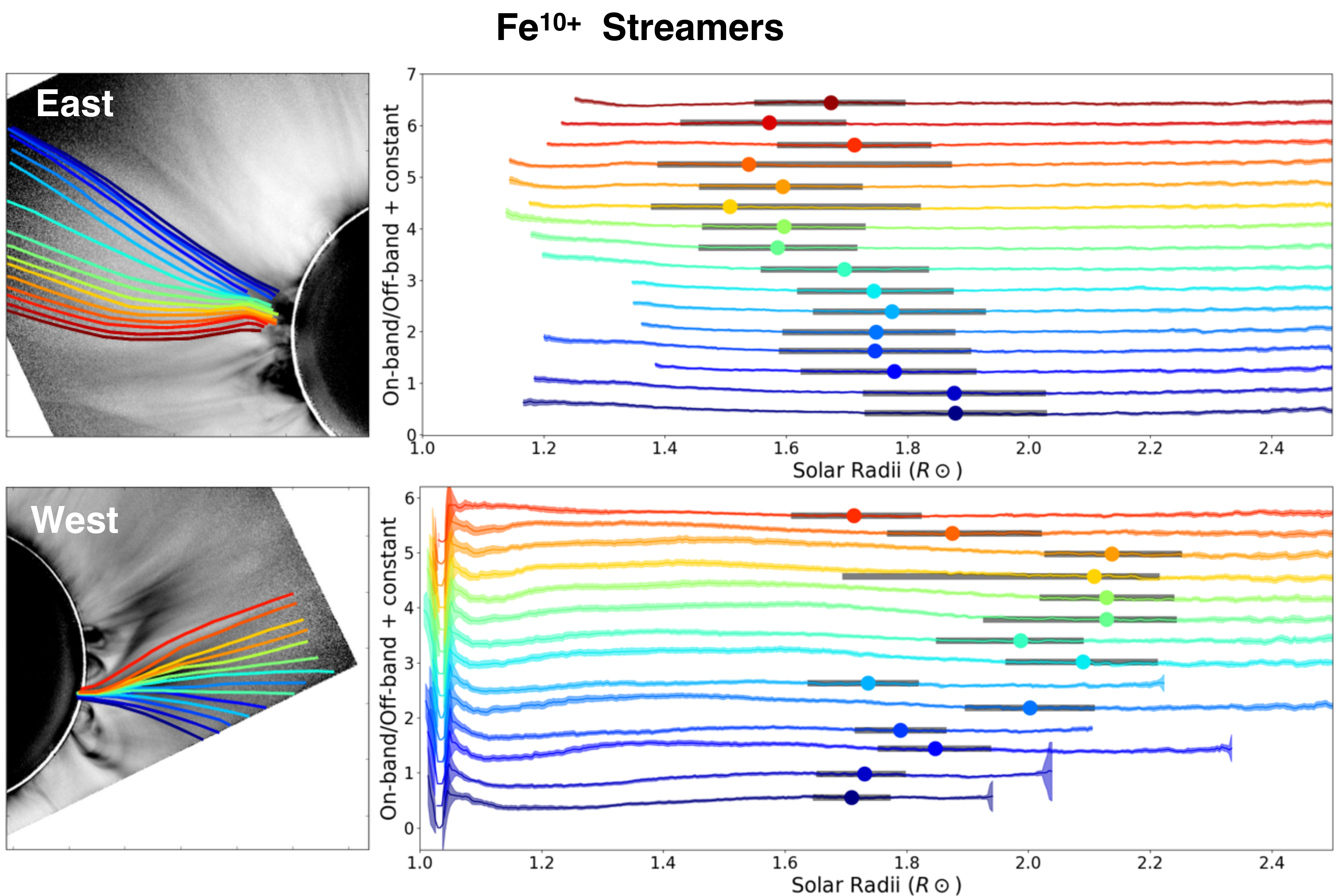}\\
\vspace*{3mm}
\includegraphics[width=6.0in]{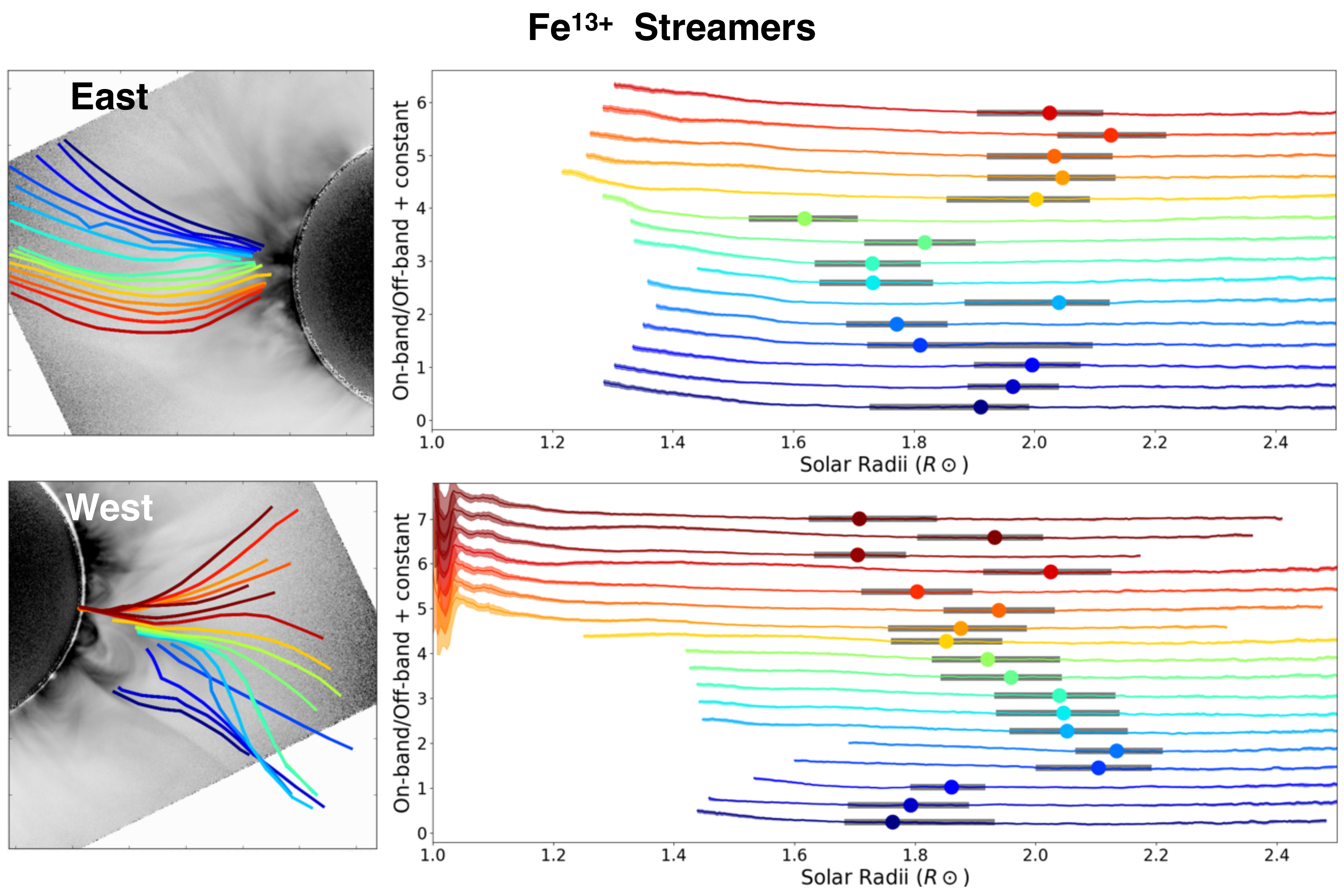}
\caption{Same as Figure \ref{fig3} for regions near the equator dominated by streamers. The top panels for each ion show an example from the east streamers, while an example from the west are on the bottom. }
\label{fig4}
\end{figure*}

\begin{figure*}[h!]
\centering
\includegraphics[width=6.25in]{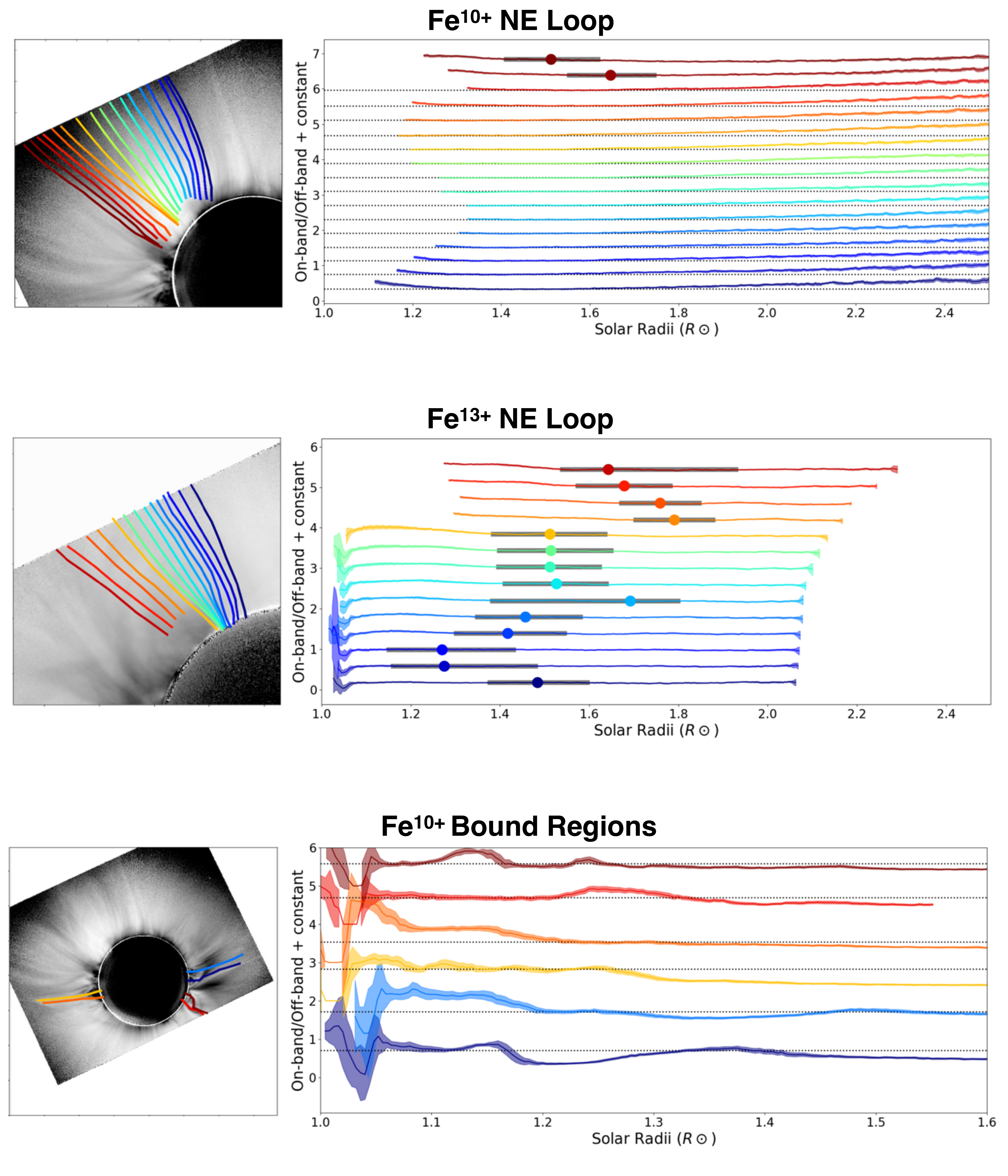}
\caption{Top: Same as Figures \ref{fig3} and \ref{fig4} for $\rm Fe^{10+}$ above the large prominence in the NE where lower limits were determined for the freeze-in distance. Note that almost every ratio profile in this region fails to freeze-in within our data. This rise is significant to over $5 \sigma$ from the original freeze-in determination before the data is noise dominated. Middle: Same as Top for $\rm Fe^{13+}$. Bottom: Variation of $f(R)$ for $\rm Fe^{10+}$ across closed loops at the base of the corona prior to reaching open field lines.}
\label{fig5}
\end{figure*}

The data numbers used for each pixel are taken as the median average from a circle of radius $r$ pixels around the pixel of interest in order to limit noise in the measurement. The scatter of pixel values in each circle is then used to determine the Gaussian $1 \sigma$ error on the median. These uncertainties determine the $\pm 1 \sigma$ ratio values to be used in the error analysis. Note that we did not determine the relative flux calibration between the images since only the slope of the ratio, and not the absolute value, is needed to determine \Rf. Flat fielding and dark frame removal are sufficient to ensure that each image is self-calibrated across the image plane.

The freeze-in distance is then determined for these curves by testing every point, $l_i$ along the line with length $L$ (in units of pixels), and checking if it meets the following conditions:
\begin{eqnarray}
\frac{\Big | g(l_i + \Delta l) - g(l_i - \Delta l) \Big |}{g(l_i)} & <  \frac{\Delta g}{g}_{max}, \\ 
\Big |\frac{\delta g}{\delta l} \Big | &< \frac{\delta g}{\delta l}_{max}, \\
l_i  &> l_{min},
\end{eqnarray}
where the $g(l_i)$ is the value of the onband/offband ratio at a given point, $\frac{\Delta g}{g_{max}}$ is the maximum allowed absolute fractional change in the ratio over the given test window ($\Delta l$) and similarly $\frac{\delta g}{\delta l}_{max}$ is the maximum allowed instantaneous gradient in the ratio profile ($\frac{\delta g}{\delta l}$). $l_{min}$ is the minimum allowed freeze-in distance for the given line which is set for each grouping of lines to prevent small flat regions from being incorrectly selected before the true freeze-in distance. 

The exact values for these terms are set by the noise and slope characteristics of a given set of lines. For all lines considered in this work, these parameters ranged from $\frac{\Delta l}{L} = 3-12\%$, $\frac{\Delta g}{g}_{max} = 0.5-3\%$ and $\frac{\delta g}{\delta l}_{max} = 0.5-3\%$ with $r=3$ pixels. Every point along the curve was tested for the freeze-in criteria, and the point at the lowest solar radius that met the criteria was taken as the location of freeze-in distance (\Rf) provided it was above $l_{min}$. The diversity of ratio profiles necessitated the use of variable freeze-in criteria to handle every freeze-in measurement properly. Polar coronal hole profiles for example tend to have a simple exponential drop with distance and remain flat after reaching the freeze-in distance. Ratio profiles within streamers follow a similar exponential drop once the structures/field lines have become open, only after clearing any closed loops. (Note that all lines were drawn from the edges of loop regions and not inside them). 
\par 
In the event that a ratio profile continues to change up to the edge of the radial extent of our data, the algorithm automatically defaults to a lower limit determination. This lower limit is used if the value of the ratio rises or falls by more than $5 \sigma$ (as determined by pixel scatter) from the originally determined $g$(\Rf) while the data remains above a signal-to-noise ratio (SNR) of 3. A lower limit measurement indicates that the ion has not yet frozen. 
\par
In the raw ratio images there is a non-physical rise in the ratio values near the edges of the image due to a noise effect. Since the offband images were already subtracted from the original ion images (to generate the onband image), the minimum noise level in the onband images is higher than that in the offband images. The signal in the ion images was much higher since it contained emission from both the ion and the electron scattering, whereas the offband images contain only electron scattering. The Poisson error of the ion images is therefore much higher (goes as $\sqrt{N}$), so when the ion images are subtracted the noise is proportionally increased compared to the signal. Consequently, the ratio of the onband and offband images will increase in a region with no coronal emission due simply to the noise floor level and not due to an actual change in the relative ion density. This effect is apparent in the center of the moon for example, where the ratio rises to the same value as seen the edge of the images (see the middle panels of Fig. \ref{fig2}). Only a rise that is statistically significant ($>5 \sigma$) and within the SNR$>3$ region of our data was considered for a lower limit on the freeze-in distance measurement. Images showing the data with SNR$>3$ are shown in the bottom panels of Figure \ref{fig2}. 

\par
\Rf\ was remeasured two more times for each curve with new pixel values coming from the median value $\pm1\sigma$ as calculated in each pixel window with radius r. The resulting two additional ratio profiles were taken as the $\pm1\sigma$ boundaries of the measured ratio profile. Error in the freeze-in distance for each line was then determined as the minimum and maximum freeze-in measured from the median and $\pm1\sigma$ ratio lines in addition to the size of the window ($\Delta l$) used for the ratio determination. 

Examples of this technique applied to the \ion[Fe xi] and \ion[Fe xiv] data are shown in Figure \ref{fig3} for the north and south coronal holes. The colored dots on each ratio profile in the right panels of these figures indicate the measured \Rf, while their associated error bars indicate the minimum and maximum values of \Rf\ as described above. Note that the dots encircled by an ellipse corresponds to \Rf\ values measured around the boundaries between the coronal hole and the streamer to its north. Similarly, examples of calculated \Rf\ values for streamers are shown in Figure \ref{fig4}, with a unique case detailed in Figure \ref{fig5}. The determination of \Rf\ in these regions had to take into consideration the existence of closed/bound loops at the base of the streamers, when present, and therefore do not always trace all the way down to the solar surface.   

\begin{figure*}[t!]
\centering
\includegraphics[width=5in]{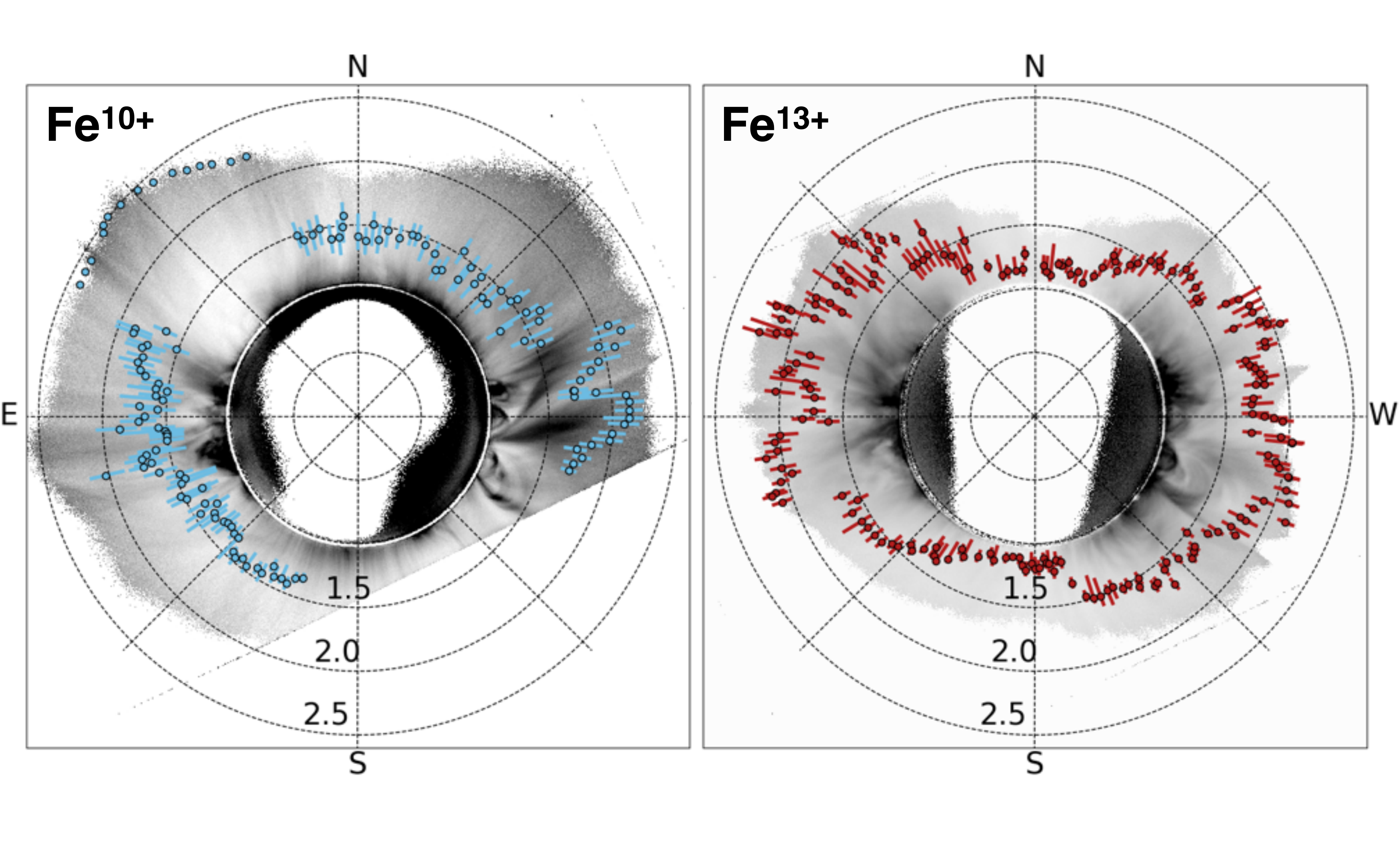}
\includegraphics[width=5in]{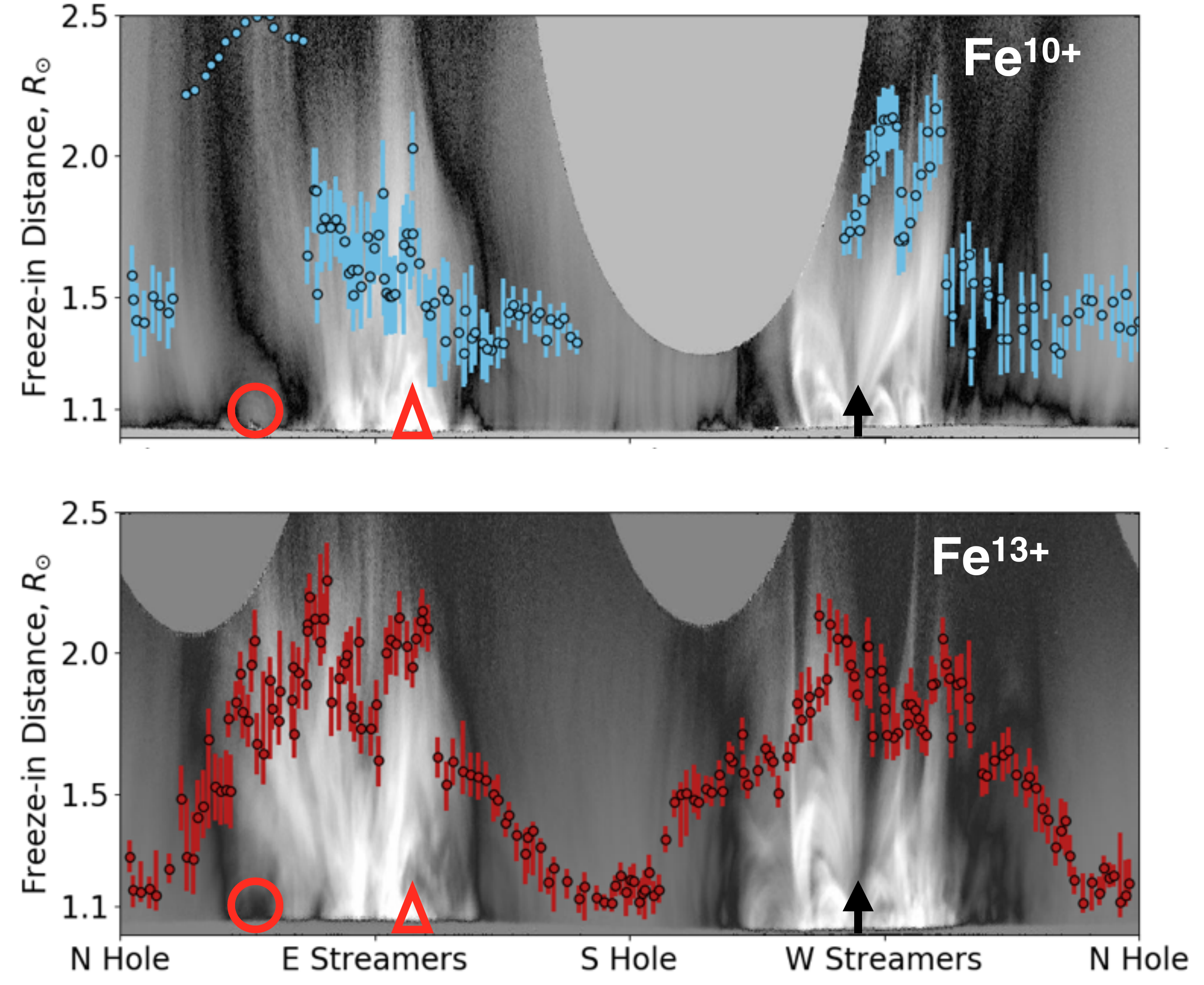}
\caption{Freeze-in distance (\Rf) measurements for $\rm Fe^{10+}$ and $\rm Fe^{13+}$ plotted over the inverted ratio on/off images in radial coordinates (top panels) and over the onband images displayed in cartesian coordinates (bottom panels). The dotted polar grids in the top panels are in steps of 0.5 $ R_\odot$.\ 
The points represent the measured \Rf\ value and the extent of the bars shows the uncertainty in the measurement as described in Section \ref{sec_FreezeIn}. Points are color-coded in blue for $\rm Fe^{10+}$ and red for $\rm Fe^{13+}$. Points without bars represent a lower limit for \Rf\ and not \Rf\ itself. The red circle and triangle point to prominences at the base of the corona. The black arrow points to the source of a CME which created a wake with its passage prior to the eclipse.}
\label{fig6}
\end{figure*}

This technique was implemented for N individual lines around the solar disk, with N = 134 for Fe$^{10^+}$ and N = 185 for Fe$^{13^+}$. The resulting collection of \Rf\ values are plotted in the top panels of Figure \ref{fig6}, where they are overlaid on the color-inverted ratio image in a polar coordinate system for each ion separately. In the bottom panels the ratios are overlaid onto their respective onband images which have been transformed into cartesian coordinates to facilitate the placement of \Rf\ values within the context of the underlying structures in coronal holes and streamers.

\section{Results}
\label{sec:res}

\subsection{Coronal holes, streamers, and streamer boundaries}

Examples of select lines along which \Rf\ was computed are given in Figures \ref{fig3} and \ref{fig4} for different coronal structures. They show how \Rf\ covers a range of values for each ion within polar coronal holes and within streamers. An example of a streamer boundary is shown encircled by an ellipse in the lower panel of Figure \ref{fig3} for $\rm Fe^{13+}$. It is clearly distinct from the polar coronal hole values for that ion just to the south. 

All \Rf\ values thus determined across the 2D images of the corona in both ions are given in Figure \ref{fig6}. The \Rf\ data are overlayed on the onband/offband ratio images in the top panel to show the relative ion density, while the lower panels use the flattened onband images (transformed to Cartesian coordinates) to show the bulk plasma density. Note that regions with enhanced density in general have a greater \Rf\ distance for both ions. The \Rf\ values clearly show the difference between the broadly defined polar coronal holes and streamers, as well as variations within each, due to the underlying fine-scale coronal structures.  While there is a sharp transition in values between coronal holes and streamers for $\rm Fe^{10+}$, a smoother transition is evident in $\rm Fe^{13+}$ as seen in Figure \ref{fig6}.

For $\rm Fe^{10+}$ we find that \Rf\ ranges from 1.25 to 1.5 \Rs\ in the polar coronal holes and 1.5 to 2.2 \Rs\ in streamers, with the exception of the north east streamer where the freeze-in distance could not be determined within the limit of the field of view of the image (see Figure \ref{fig5}). $\rm Fe^{13+}$ on the other hand has some \Rf\ values below 1.2 \Rs\ within the polar coronal holes, which are the lowest values observed in this study. The transition between the coronal holes and streamers is clearly evident in the \Rf\ values as they rise from 1.2 to 1.6 \Rs, starting from the coronal holes towards the streamers. \Rf\ varies between $\approx$ 1.7 and 2.25 \Rs\ within the streamers. 

\begin{table}[h]
\begin{center}
{\sc \Rf \ in coronal holes and streamers}
\begin{tabular}{ c  c  c  c  c }
\hline
Ion & N. coronal hole &  S. coronal hole &  E. streamers &  W. streamers \\
\hline
$\rm Fe^{10+}$ & 1.44 $\pm$ 0.06 & 1.42 $\pm$ 0.05 & 1.59 $\pm$ 0.17& 1.71 $\pm$ 0.26 \\
\hline
$\rm Fe^{13+}$ &1.19 $\pm$ 0.08 & 1.22 $\pm$ 0.14 & 1.82 $\pm$ 0.24 & 1.74 $\pm$ 0.18\\
\hline
\label{table1}
\end{tabular} 
\end{center}
\vspace{-9mm}
\caption{\sc \small Median \Rf\ in units of \Rs\ for $\rm Fe^{10+}$ and $\rm Fe^{13+}$ for coronal holes and streamers with $\pm 1 \sigma$ Gaussian scatter for each region. \label{table1}}
\end{table}

The \Rf\ measurements are separated broadly into North and South polar coronal holes as well as East and West equatorial streamers in Table \ref{table1}. The values represent the median \Rf\ in each region for $\rm Fe^{10+}$
and $\rm Fe^{13+}$ together with their corresponding $1\sigma$ Gaussian error on the scatter of \Rf. The east and west streamer regions are referred to as `streamers' in Table \ref{table1} since they have a complex structure and do not necessarily represent a single streamer. Excluded from this table is the northeast region where $\rm Fe^{10+}$ had not frozen in by the edge of our field of view, which will be discussed separately.

Inspection of Table \ref{table1} shows that there is no measurable difference in the median values of \Rf\ between the two polar coronal holes for each ion. Although the extent of the south polar coronal hole beyond 1.25 \Rs\ was cutoff in the southwest Fe$^{10+}$ data (see Figure \ref{fig1}D), the fact that the freeze-in distance for $\rm Fe^{10+}$ could not be computed below the cutoff in the data is an indication that its \Rf\ there could not be smaller than \Rf\ for $\rm Fe^{13+}$. Had it been closer to the Sun, we would have been able to determine it from the available radial extent of the data. An \Rf\ distance above the lower limit in the southwest is consistent with the \Rf\ values for Fe$^{10^+}$ seen in the north coronal hole. 

The \Rf\ values for $\rm Fe^{10+}$ are significantly different between the East and West streamers. Its value above the west streamers is comparable to that of $\rm Fe^{13+}$ there, which is consistent with the mixture of \ion[Fe xi] and \ion[Fe xiv] emission seen in Figure \ref{fig1}C in that region. Above the east streamers the \Rf\ for $\rm Fe^{10+}$ is significantly smaller than for $\rm Fe^{13+}$. As a matter of fact, it is within $1\sigma$ of its value in either coronal holes. 

The presence of cool material within the west streamer region is also evident in the dark wedges seen in the \ion[Fe xiv] image in comparison to the eastern region which has much more continuous \ion[Fe xiv] emission near the equator. This wedge is located at the site of a CME which erupted about ten hours prior to the eclipse observations. The impact of CMEs will be discussed in a section to follow.

\begin{figure*}[t!]
\centering
\includegraphics[width=4.4in]{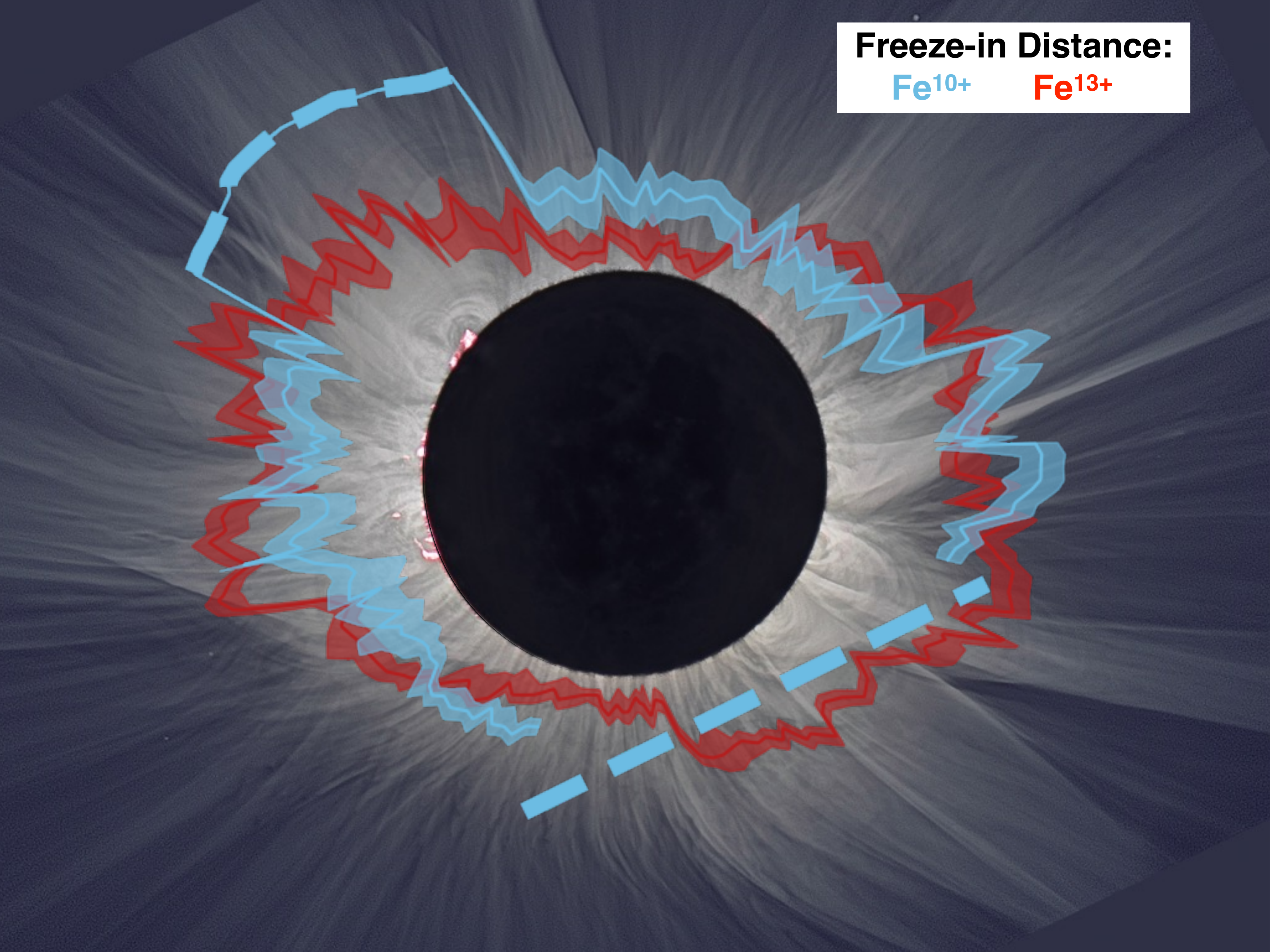}
\vspace{5mm}
\includegraphics[width=4.4in]{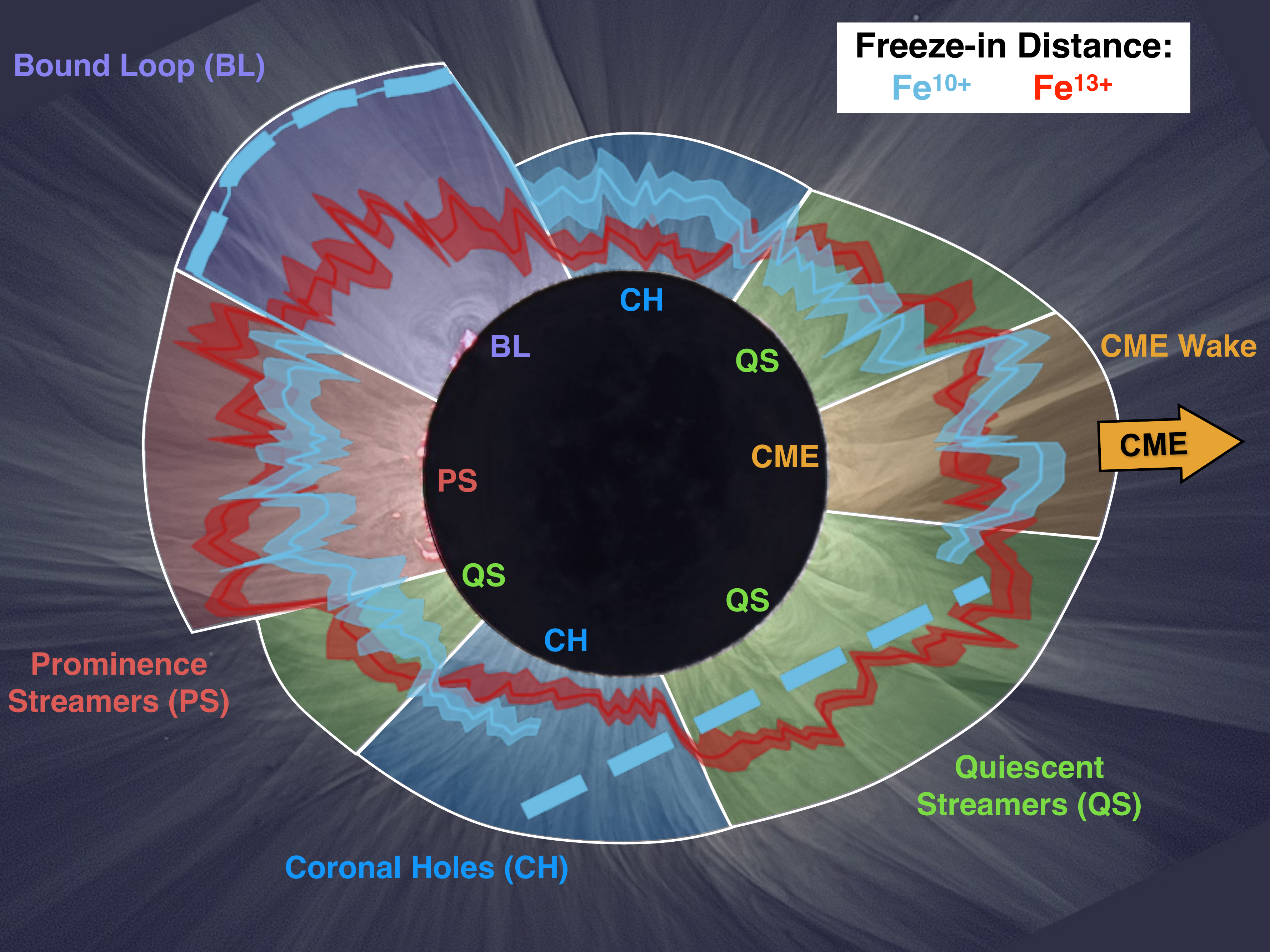}
\caption{Top: Overlay of Fe$^{10+}$ (blue) and Fe$^{13+}$ (red) freeze-in distance measurements on the white light image eclipse image. The central bold line indicates the interpolated measurements of the freeze-in distances and the thickness of the transparent band indicates the interpolated confidence limit of the measurement determined by the technique demonstrated in Figures \ref{fig3} and \ref{fig4}. The dashed blue line indicates the edge of the \ion[Fe xi] emission image representing a lower limit on the freeze-in distance in that region. Bottom: Same as above but with labels indicating different regions in the corona as identified by the combination of freeze-in distances. Note that the freeze-in distances of \ion[Fe xiv] in the Bound Loop region are not necessarily correct given the dynamics of \ion[Fe xi] (See Fig. \ref{fig5}).}
\label{fig7}
\end{figure*}

A visual distinction of the differences in \Rf\ values between the two ions is given in Figure \ref{fig7} with the blue Fe$^{10+}$ and red Fe$^{13+}$ bands representing their corresponding \Rf\  values. The bands are superimposed on the high resolution white light image eclipse image (from Figure \ref{fig1}) to put the relative \Rf\ values in the context of coronal structures. The central bold line indicates the interpolated measurements of the freeze-in distances and the thickness of the transparent band indicates the interpolated confidence limit of the measurement determined by the technique demonstrated in Figures \ref{fig3} and \ref{fig4}, and described in Section \ref{sec_FreezeIn}. The dashed blue line indicates the edge of the \ion[Fe xi] ion image representing a lower limit on the freeze-in distance in that region. The same overlay is given in the lower panel, with labels, e.g. coronal holes (CH) and quiescent streamers (QS) indicating different regions in the corona as identified by the combination of freeze-in distances of the two ions. Table \ref{table2} summarizes these differences. Details pertaining to the impact of prominences and CMEs are described next.

\begin{table}[h]
\begin{center}
\begin{tabular}{c }
{\sc \Rf \ by coronal morphology}
\end{tabular} 
\begin{tabular}{ c c  c  c  c }  
\hline
Ion & Coronal Holes &  Quiescent Streamers &  Prominence Streamers & CME Wake  \\
\hline
$\rm Fe^{10+}$ & 1.44 $\pm$ 0.06 & 1.45 $\pm$ 0.14 & 1.65 $\pm$ 0.14 & 1.99 $\pm$ 0.16 \\
\hline
$\rm Fe^{13+}$ &1.19 $\pm$ 0.09 & 1.62 $\pm$ 0.18 & 2.03 $\pm$ 0.17 & 1.81 $\pm$ 0.11\\
\hline
\end{tabular} 
\end{center}
\vspace{-4mm}
\caption{\sc \small Median \Rf\ in units of \Rs\ for $\rm Fe^{10+}$ and $\rm Fe^{13+}$ for regions shown in Figure \ref{fig7} with $\pm 1 \sigma$ Gaussian scatter for each region \label{table2}}
\end{table}

\subsection{Impact of prominences}

Prominences are invariably localized at the base of streamers as reported from eclipse observations by \cite{Habbal2010}. Indeed, there are prominences visible in Figure \ref{fig1}A down a section of the east limb with bound loops and streamers above. It is clear from Figure \ref{fig7} that there are distinct prominences above which \Rf\ values for Fe$^{13^+}$ are significantly higher than in the surrounding streamers with smaller prominences. These regions will hereafter be referred to as `Prominence Streamers' (see Table \ref{table2}). The \Rf\ values in this region for $\rm Fe^{13+}$ range between $\approx$ 1.75 and 2.25 \Rs, which is higher than anywhere else in the corona. The \Rf\ for $\rm Fe^{10+}$ on the other hand is more complex. In the prominence streamers $\rm Fe^{10+}$ freezes-in only slightly beyond its values in the Quiescent Streamer regions.

Figure \ref{fig5} displays a peculiar case above a prominence where Fe$^{10^+}$ did not freeze-in within the radial extent of our data. This region happens to be located directly above the largest prominence visible during the eclipse. The observed ratio profiles decrease to a flat minimum as normal, but then increase significantly past 2 \Rs. Careful inspection of the white light image in Figure \ref{fig1}A reveals that there are loops extending to the edge of our onband data in this region. The plasma here is therefore bound and cannot freeze-in within the limits of our data. It is interesting to note that Fe$^{13^+}$ shows no evidence of change beyond the calculated \Rf\ distance as shown in the middle panel of Figure \ref{fig5}. The bottom panel of the figure shows an example of ratio profiles drawn through bound regions in the \ion[Fe xi] image; where the ratio profiles fluctuate a great deal while inside the bound loops, then transition to a simple exponential drop and flatten once the field lines become open. If we had narrowband data extending to well beyond the top of this large bound loop, it is conceivable that we would see the same effect as in other bound regions and could measure the freeze-in distance above the bound loop once the field lines become open, perhaps at 3-5 \Rs. Thus this region is labeled as `Bound Loop' (BL) in Figure \ref{fig7} and the \Rf\ values in this region are ignored for calculation of values in Table \ref{table2}.

\subsection{Impact of CMEs}

A CME was observed by SDO and SOHO near the plane of sky in the west just south of the equator about 10 hours prior to the eclipse, as given by the arrow in Figure \ref{fig6}. This CME left an open wedge in the corona for hours after its passage, which is especially apparent in the absence of \ion[Fe xiv] emission in that image (see Figure \ref{fig2}). However, \ion[Fe xi] emission is still present in this wake indicating that the temperature is significantly lower than the surrounding regions as a result of the CME passage. This CME wake is the only region in the corona where \Rf\ for $\rm Fe^{10+}$ is larger than \Rf\ for $\rm Fe^{13+}$, other than in coronal holes. This wake is labeled as `CME Wake' in both Figure \ref{fig7} and the corresponding \Rf\ values in Table \ref{table2}. Note that the CME wake had a small streamer inside toward the North (see Fe$^{10+}$ image in Fig. \ref{fig6}) which changed the freeze-in distances to be similar to that of the Quiescent Streamers. The single northern streamer is projected in the same line of sight as the remaining CME wake, and so the freeze-in distance inferred is an average of the two regions and so is left as part of the `CME Wake' in Table \ref{table2} and Figure \ref{fig7}. Streamers toward the south also were caught in the CME wake but had fully reverted to typical QS behaviour by the time of the eclipse several hours afterward, and so are grouped with the other Quiescent streamers.

A second CME had originated in the corona prior to the eclipse observations behind the plane of sky and was observed with spectral data (see \citealt{Ding2017}). The Doppler redshifts of filamentary material forming it were equivalent to speeds of 1000-1500 km s$^{-1}$ at the time of the eclipse. Unlike the plane of sky CME observed in the west, this behind the plane of sky CME has no discernible effect on the freeze-in distance measurements. A comparison of the two CMEs offers proof that the emission from the \ion[Fe xi] and \ion[Fe xiv] lines, captured in these images, is largely confined to the plane of sky.  

\section{Discussion}
\label{sec:disc}

In general we find that Fe$^{10^+}$ and Fe$^{13^+}$ have variable freeze-in distances throughout the corona. These two ions typically freeze-in at different heliospheric distances along the same line of sight indicating either that the different ions are dominant along different field lines (which are projected into the same line of sight), or that intermediate ions (Fe$^{11^+}$ and Fe$^{12^+}$) are enabling either Fe$^{10^+}$ or Fe$^{13^+}$ to continue charge state evolution after the other has frozen in. However, the ionization and dielectronic recombination rates are such that if the ions emit from the same plasma tube, then when the lower charge state (say Fe$^{9^+}$) freezes in, then so should the subsequent higher charge states (i.e. Fe$^{13^+}$), and the freeze-in distances should be identical. The fact that we infer different freeze-in distances suggests that the two ions belong to different plasma flow tubes. Freeze-in distance observations of additional ions are required in order to further explore these differences.

As shown in the labeling of the lower panel of Figure \ref{fig7}, changes in \Rf\ for the two ions across different structures can be resolved into four main regions: Coronal Holes, Quiescent Streamers, Prominence Streamers and a CME wake. In the coronal holes (CH) where Fe$^{10^+}$ consistently freezes-in at about 1.5 \Rs, Fe$^{13^+}$ has a lower freeze-in distance around 1.2 \Rs. 
The low density is such that the plasma is unable to either ionize Fe$^{12^+}$ into Fe$^{13^+}$, and too little Fe$^{14^+}$ is likely present to recombine into Fe$^{13^+}$, thus leading the ion to freeze immediately above the solar surface. By contrast, the low temperature enables Fe$^{10^+}$ to continue to evolve out to 1.5 \Rs. 

This scheme is reversed in the typical streamers, which we refer to as `Quiescent Streamers' (QS), where Fe$^{10^+}$ and Fe$^{13^+}$ trade places in the order of their freeze-in distances compared to the holes. In the lower parts of the streamers where the magnetic field lines are still largely bound to the Sun, the inference of a freeze-in distance is meaningless. Above these closed arches however, the field lines are open and ions can escape with the solar wind. This is demonstrated by the inferences of the freeze-in distances in the streamers where the ratio profiles fluctuate a great deal inside the bound loops (see Figure \ref{fig5}). However, they have a similar behavior to coronal holes once the field lines appear open. The bound structures therefore raise the escape distance of the solar wind compared to coronal holes, but do not prevent the solar wind from forming above them. In the Quiescent Streamers, Fe$^{10^+}$ continues to freeze-in at about 1.4-1.6 \Rs\ (like the coronal holes) while Fe$^{13^+}$ freezes in around 1.5-1.7 \Rs. The exact values of \Rf\ fluctuate slightly from region to region, but \Rf\ for Fe$^{13^+}$ is consistently larger than for Fe$^{10^+}$. These regions therefore are likely to have a significantly higher temperature, thus allowing Fe$^{13^+}$ to be created. The higher but more variable electron density leads to fluctuations in \Rf\ for both ions. 

It is interesting to note that at the boundary of the southern coronal hole and the western streamers there is an increase in the freeze-in distance for Fe$^{13^+}$ (see Fig. \ref{fig3}). This rise is not entirely unexpected, as the boundaries of coronal holes have been thought to be a possible source for the slow solar wind (e.g. \citealt{Sakao2007, Antiochos2011, Riley2012, Stakhiv2016}). If this is the case, we would expect the boundaries to have a higher density (and slower flow speed) than the center of the hole which would increase the freeze-in distance at the boundary, as we have observed. Furthermore, the even larger freeze-in distances observed elsewhere in the corona support the idea that different coronal morphologies create a plethora of different solar wind types and that no single source is sufficient to explain the slow solar wind (see \citealt{Xu2015}). 
\par

The behaviour at the coronal hole boundary could also be explained as a line of sight projection effect. It is possible that streamers slightly off the plane of sky are contaminating the coronal hole signature near the boundary. Ambiguity in the line of sight observation is a recurring issue in all remote sensing observations of the corona and could be altering the freeze-in distance measurement. The slow transition of freeze-in distances therefore would represent a smooth change in the average structure along the line of sight, from hole to streamer. Another example of this ambiguity is the case of the CME wedge with the small streamer inside it. In the streamer we measure a freeze-in distance typical of other quiescent streamers for both ions, while the rest of the wedge is dominated by more typical coronal hole-like behavior. Even though there may be ambiguity to the exact structure that is being sampled, we can reasonably conclude that whichever coronal morphology is dominant near the plane of the sky will dominate the measurement of freeze-in distance based on the totality of our observations.

\par
Prominences have been observed to cause large scale turbulence in the corona right above them \citep{Druckmuller2014}. They have also been shown to be intricately linked to large scale structures seen in white light \citep{Habbal2014, Druckmuller2017} and to be enshrouded by the hottest material in the corona \citep{Habbal2010}. Hence, it is not totally surprising that prominences could also impact the freeze-in distance as discovered in the \Rf\ inferences in the `Prominence Streamer' (PS) region. The presence of turbulence around prominences, and their intricate connection with the large scale coronal structures, can thus account for this freeze-in behavior if the resulting turbulence is acting to heat the plasma beyond the peak ionization of Fe$^{10^+}$ and closer to that of Fe$^{13^+}$. In this case the Fe$^{10^+}$ ion density will be depleted by ionization, while the Fe$^{13^+}$ density becomes enhanced by ionizing lower states causing the freeze-in Fe$^{13^+}$ distance to increase as seen in Figure \ref{fig7}. Support for these arguments can also be found in the ratio images (see Figure \ref{fig2}) which show that Fe$^{10^+}$ has a higher ion density in the west where there are no prominences compared to the east, and vice versa for Fe$^{13^+}$ where the ion density is higher in the east above the prominences. The Quiescent Streamers are an interesting case between the extremes of the Coronal Holes and the Prominence Streamers. The Quiescent Streamers have a larger temperature and density than the Coronal Holes causing a rise in the Fe$^{13^+}$ \Rf\ distance, yet they have much smaller bound loops and lower turbulence in comparison to the Prominence Streamers causing both ions to have a comparatively lower QS freeze-in distance. 

In the north east prominence streamer the ratio of Fe$^{10^+}$ rises significantly after an initial minimum is reached indicating that the ion has not yet frozen in within our data. The white light image shows some bound field lines out to the limit of our narrowband data indicating that this effect may be due to bound coronal plasma which cannot freeze-in in the northeast. In this case we are observing the behavior of bound plasma which is kept at a high temperature near the turbulent prominences and cools near the top of the 2 \Rs\ sized loops where Fe$^{10^+}$ recombines causing the ratio to rise again as seen in Figure \ref{fig5}. It is possible that the Fe$^{13^+}$ has also not fully frozen-in by the edge of our data either in this northeast prominence streamer, but this is not possible to measure as the ratio remains sufficiently flat to the edge of our high SNR data. Note that the NE region of the \ion[Fe xi] data is the only place where any ratio profiles rose with statistical significance inside the SNR$>3$ data. Eventually the ratio rises everywhere at the edge of the image when the signal fades due to a noise effect  (see Section \ref{sec_FreezeIn} and Figure \ref{fig2}).

\par

The same relative difference in \Rf\ between the two ions was observed in the wake caused by the passage of a CME, albeit with larger values for both. This could imply that the passage of the CME likely reduced the electron temperature and density causing Fe$^{10^+}$ to dominate in the same manor as in the coronal holes. The lower temperature can be inferred by the increased \ion[Fe xi] emission in the wake (see Fig. \ref{fig1}C) in comparison to \ion[Fe xiv] emission, especially in contrast to the nearby streamers. However, the region is surrounded by high density streamers which likely maintains a higher electron density than the coronal holes causing both ions to freeze-in at a much larger distance.

\section{Conclusions}
\label{sec:conc}
In this paper we have presented the first observational inference of the freeze-in distance for $\rm Fe^{10+}$ and $\rm Fe^{13+}$ in the solar corona using data from the 2015 March 20 total solar eclipse.  We find that \Rf 's inferred for each ion are different, being closest to each other in polar coronal holes and highest above prominences and in the wake of a CME. In general we find that the solar wind originating from polar coronal holes has a smaller \Rf\ than in typical quiescent equatorial streamers (see Figs \ref{fig6} and \ref{fig7}). 

The complexity of plasma structures throughout the corona, derived from these eclipse observations show how the derived contours of freeze-in distances would have been practically impossible to infer from in-situ measurements alone, since they are spatially limited to much smaller volumes, projecting back to a small fraction of the coronal structures at any given time. Models would have also fallen short in accounting for this determination, as the plasma conditions in-situ cannot be inferred with the spatial resolution available in the imaging data. 

\par
Several pieces of information are surprisingly captured in the eclipse images given that they are snapshots of the instantaneous state of the corona. (1) The impact of dynamic events in the corona, such as the imprint or impact of the passage of a CMEs and presence of prominences, is captured in the freeze-in distance; (2) The diversity of coronal structures on spatial scales of a few arc seconds translate into the same spatial resolution in the freeze-in distance indicating large variations in the plasma conditions along different field lines. While little information regarding the solar wind speed can be gleaned from the eclipse observations, it is reasonable to conclude that the small scale variations in the freeze-in distances between the two ions considered here are indications either of differential flows or differential electron density profiles (see \citealt{Ko1997}). However \cite{Owocki1983a} found that the electron temperature had the largest effect on the freeze-in distance which would mean that the difference in freeze-in distances is a result of varying temperatures rather than flow speeds or electron density. It is quite possible that all three of these physical conditions are affecting the final freeze-in distance. 
\par
Prior to the empirical inference of \Rf's presented in this work, estimates of \Rf's for different ions were undertaken with empirical model studies using primarily in-situ measurements 
in combination with observations of the electron density in the inner corona. Their calculated value of \Rf\ was then directly linked to the electron temperature in the corona.
It is now clear that no single temperature can account for the spatial distribution of the \ion[Fe xi] and \ion[Fe xiv] emission (Fig. \ref{fig1}) and the diversity of \Rf\ values for the two ions (see Fig. \ref{fig7}). The close proximity and somewhat constant nature of the contours in the polar coronal hole regions are the best indication that there could be a single freeze-in temperature there. However, that is far from being the case in the other regions of the corona where the difference between \Rf\ for Fe$^{10^+}$ and Fe$^{13^+}$ is very pronounced. Our observations support the modeling results of \cite{Burgi1987}, \cite{Esser1998} and \cite{Landi2012b} that the electron temperature inferred from charge state measurements strongly depends on model assumptions (as evidenced by the large variations in freeze-in distance), and that ``none of the theoretical models are able to reproduce all observations; namely all of them underestimate the charge state distribution of the solar wind" \citep{Landi2014}. This result indicates that averaging in-situ charge state measurements over long time periods can yield misleading results \citep{Landi2012a}. Furthermore, the results presented here show that models based on the assumption of a constant source surface with a fixed solar radius are misleading given the large range of inferred \Rf\ values for both ions. 

Our observations also underscore the inherent difficulty in deriving plasma parameters in the inner corona solely from in-situ charge state measurements, and/or EUV observations. In-situ data alone cannot provide the detailed determination of the freeze-in distance in two-dimensions, because these observations are inherently limited to the small volume of space probed by the detectors in space, and because they rely on empirical models. On the other hand, plasma diagnostics based on EUV data rely on the fact that the plasma is in ionization equilibrium within the field of view of these imagers. However, the results presented here show that in coronal holes, the plasma is often frozen in at distances well within the field of view of current EUV instruments (such as AIA, EIS, and XRT), i.e. where ionization equilibrium is not valid. Consequently, most of the standard diagnostic techniques based on EUV data cannot be applied uniformily throughout the corona.

%

\subsection*{Acknowledgments}
We thank Judd Johnson, Scott Gregoire and Peter Aniol for their outstanding technical assistance throughout the preparation for the eclispse and the observations during totality.  These eclipse observations would not have materialized without the local arrangements provided by Prof. Fred Siegernes and his staff at UNIS, as well as by Mr. Morten Ulsnes, manager of the Svalbard airport on Longyearbyen. Financial support was provided by NASA and NSF under NASA Grant NNX13AG11G and NSF Grants AGS-1144913, AGS-1358239, and AGS-1255894, to the Institute for Astronomy of the University of Hawaii, and
NASA grant NNX15AB73G and NSF grants AGS-1255704 and AGS-1408789 to the University of Michigan (EL). M. Druckmüller was supported by Grant Agency of Brno
University of Technology, project No. FSI-S-14-2290. 

\bibliographystyle{apj}
\bibliography{msArxiv}

\end{document}